\documentclass[a4paper,twocolumn,11pt,unpublished=2017-05-09]{quantumarticle}

\pdfoutput=1
\usepackage[utf8]{inputenc}
\usepackage[english]{babel}
\usepackage[T1]{fontenc}
\usepackage{amsmath}
\usepackage{amsfonts}
\usepackage{bm}
\usepackage{tikz}
\usepackage{hyperref}
\usepackage{dsfont}

\usepackage{tikz}
\usepackage{lipsum}
\PassOptionsToPackage{compress}{natbib}
\usepackage[numbers]{natbib}
\usepackage{braket}

\usepackage{optidef}
\usepackage{mathtools}
\DeclarePairedDelimiterX{\inner}[2]{\langle}{\rangle}{#1, #2}

\usepackage[compat=0.4]{yquant}
\makeatletter

\DeclareRobustCommand{\rvdots}{%
	\vbox{
		\baselineskip4\p@\lineskiplimit\z@
		\kern-\p@
		\hbox{.}\hbox{.}\hbox{.}
}}
\makeatother
\yquantdefinebox{dots}[inner sep=0pt]{$\rvdots$}
\yquantdefinebox{hdots}[inner sep=0pt]{$\ldots$}
\usepackage{mathtools}

\DeclareMathAlphabet{\mathbbb}{U}{bbold}{m}{n}

\newcommand{\ketbra}[1]{\ket{#1}\bra{#1}}
\DeclarePairedDelimiter\abs{\lvert}{\rvert}
\DeclarePairedDelimiter\norm{\lVert}{\rVert}

\begin{document}

\title{Provable advantages of kernel-based quantum learners and quantum preprocessing based on Grover's algorithm}
\author[1,2]{T. Muser\textsuperscript{$\dag$} }

\author[1]{E. Zapusek\textsuperscript{$\dag$} }
\orcid{0000-0002-6421-6308}
\author[3]{V. Belis}
\orcid{0000-0001-5920-8998}
\author[1]{F. Reiter}
\orcid{0000-0002-5217-3064}

\affil[1]{Institute for Quantum Electronics, ETH Z\"urich, Otto-Stern-Weg 1, 8093 Z\"urich, Switzerland}
\affil[2]{Current address: Swiss Data Science Center, ETH Z\"urich, Andreasstrasse 5, 8050 Zürich}
\affil[3]{Institute for Particle Physics and Astrophysics, ETH Z\"urich, Otto-Stern-Weg 5
8093 Z\"urich, Switzerland}

\maketitle

\begin{abstract}
There is an ongoing effort to find quantum speedups for learning problems. Recently, [Y. Liu et al., Nat. Phys. \textbf{17}, 1013--1017 (2021)] have proven an exponential speedup for quantum support vector machines by leveraging the speedup of Shor's algorithm. We expand upon this result and identify a speedup utilizing Grover's algorithm in the kernel of a support vector machine. To show the practicality of the kernel structure we apply it to a problem related to pattern matching, providing a practical yet provable advantage. Moreover, we show that combining quantum computation in a preprocessing step with classical methods for classification further improves classifier performance.

\end{abstract}
\let\thefootnote\relax\footnotetext{\textsuperscript{$\dag$}  These authors contributed equally to this work.}
\begin{figure*}[t]
    \centering
    \includegraphics[width=\textwidth]{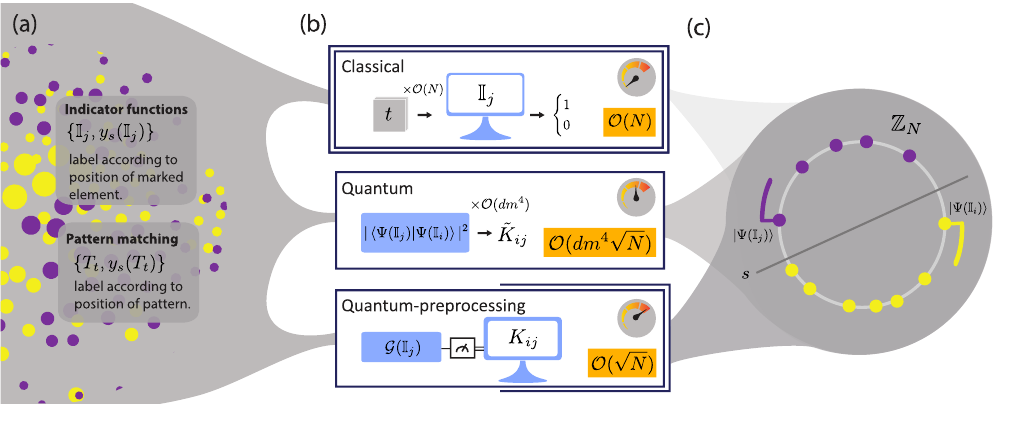}
    \caption{(a) Description of the learning problem.
    We consider two distinct problems: Classifying black-box indicator functions based on the element they represent, and classifying strings or images based on a pattern's position.     
    The approaches we apply are similar for both problems. Here we discuss the former problem in more detail. (b) Learning approaches. During inference the learner is presented with an unseen indicator function $\mathbb{I}_j$. A classical learner has to resort to brute-force sampling from the indicator function requiring $\mathcal{O}(N)$ queries. A quantum learner can use the quantum kernel, evaluating the inner product with $d$ support vectors, thereby reducing the complexity to $\mathcal{O}(\sqrt{N})$. However, to ensure noise robustness one has to sample $\mathcal{O}(m^4)$ times, where $m$ the size of the training set. Lastly, when the quantum processing is shifted to a preprocessing there is no noise robustness overhead. The output of the quantum circuit $\mathcal{G}$ can be evaluated in a single run. Moreover, classical intermediate states can be stored and accessed later, eliminating the factor $d$. Consequently, the quantum preprocessing approach exhibits the lowest sample complexity. (c) Cyclic classification problem. The classification problem is a binary classification on the cyclic group $\mathbb{Z}_N$. The labels are assigned according to the value $s$.}
    \label{fig:summary}
\end{figure*}
Quantum computing has evolved from a theoretical concept to a physical reality, with experiments that demonstrate an advantage over classical computation~\cite{arute_quantum_2019, zhong_quantum_2020, madsen_quantum_2022}. The decades long effort that led to these demonstrations is justified and driven by the performance guarantees of quantum algorithms such as Shor's and Grover's, which provide speedups for the problems of prime factorization and unstructured search~\cite{shor_polynomial-time_1997, grover_fast_1996}.

The bustling subfield quantum machine learning promises to be one of the first applications of quantum computation. Leveraging algorithms with lower performance requirements, this domain presents exciting possibilities even with current quantum devices~\cite{daley_practical_2022}. It is motivated by  the tremendous success of classical machine learning, the potential power of quantum computers~\cite{benedetti_parameterized_2019,schuld_quantum_2019,du_learnability_2021}, and promising theoretical studies and guarantees~\cite{abbas_power_2021, Huang_2021, Goto2021, Caro2022, Huang_2022, huang2021_advantage_bounds, Wu2023quantumphase}.
However, for classical data the prospect of a quantum advantage with shallow quantum circuits is still unclear~\cite{Cerezo2022}.
On the other hand, for some quantum machine learning algorithms which rely on fault-tolerant quantum computers a quantum speedup has been proven~\cite{servedio_equivalences_2004, Rebentrost2014, liu_rigorous_2021, gyurik_establishing_2022,pirnay_super-polynomial_2022}.
In particular, in Ref.~\cite{liu_rigorous_2021}, the authors Y. Liu, S. Arunachalam, and K. Temme have proven a quantum-classical separation for a quantum support vector machine (QSVM) by leveraging the speedup of Shor's factoring algorithm~\cite{liu_rigorous_2021}. The proof is built on the widely held belief that there is no efficient classical algorithm for factorization. Just as Shor's and Grover's algorithms are vital for the field of quantum computing, algorithms with provable advantages are essential to the field of quantum machine learning as their quantum speedups provide theoretical justification for the effort invested in the field. They provide an example of a quantum advantage and thereby show promise for the existence of further quantum advantages~\cite{havlicek_supervised_2019, Rebentrost2014, Huang_2021, huang2021_advantage_bounds}.

With our present work, we follow a related approach and use Grover's algorithm to construct a QSVM. Grover's algorithm provides a quadratic speedup in query complexity for the problem of unstructured search. Discovered in 1996, it is one of the most prominent quantum algorithms, with applications in Boolean satisfiability~\cite{cook_complexity_1971-1,miller_reducibility_1972,yang_solution_2009-1}, pattern matching, optimization~\cite{durr_quantum_1996,bulger_implementing_2003,baritompa_grovers_2005} and quadratic unconstrained binary optimization problems~\cite{gilliam_grover_2021}. Importantly, its speedup is provable~\cite{bennett_strengths_1997}, meaning that it can be rigorously shown that the best classical algorithm is quadratically slower than Grover's algorithm. 
 
There is a strong connection between quantum machine learning and Grover's algorithm~\cite{anguita_quantum_2003,aimeur_quantum_2013,paparo_quantum_2014,wiebe_quantum_2014,NIPS2016_d47268e9,du_grover-search_2021,salmon_provable_2023}. Many quantum machine learning algorithms use Grover's for the computationally expensive optimization during training. Models that can incorporate Grover's algorithm include support vector machines~\cite{anguita_quantum_2003}, perceptrons~\cite{wiebe_quantum_2014}, clustering algorithms~\cite{qkmeans_noisy, wozniak_belis_puljak2023}, and reinforcement learning~\cite{paparo_quantum_2014}.

In our study, we use Grover's algorithm in the kernel of a QSVM, allowing us to prove a quadratic speedup in a learning problem involving classical data (Fig.~\ref{fig:summary}(a)).
We demonstrate the practical applications in App.~\ref{app:pattern} where we show that a kernel with the same structure can be applied to a learning problem of practical relevance; classifying texts or images.

Finally, we improve on the quantum kernel learner by instead supplying a classical model with data that has been preprocessed on a quantum machine. This allows intermediate states to be stored and removes the overhead required for noise-robustness thereby reducing the complexity. The same preprocessing approach can also be applied to improve the algorithm from Ref.\;\cite{liu_rigorous_2021}.
\hfill

\section{Quantum support vector machines}\label{sec:QSVM}

Support Vector Machines (SVMs)~\cite{svmVapnik, Cortes1995, steinwart_2008} are binary, linear classifiers. Provided with a set of training datapoints, each belonging to one of two classes, the SVM constructs a model that assigns class labels to unseen datapoints. For a given set of training examples an SVM deterministically finds a hyperplane that separates the datapoints with the maximal margin.
To train an SVM, one needs to evaluate pairwise inner products of the training data $\bm{x}_i \in \mathcal{X}$. Classification of a new datum similarly requires the evaluation of inner products of the new datum with a subset of the training data, called \emph{support vectors}. Non-linear classification can be performed efficiently via the kernel trick~\cite{schoelkopf2009, theodoridis_koutroumbas_2009}: We substitute the inner product with a function $k$, known as the \textit{kernel}. This implicitly maps data with the \textit{feature map} $\varphi$ to an inner product of a different, high-dimensional Hilbert space $\mathcal{H}$, referred to as the latent or feature space;
\begin{equation}
	k(\bm{x}_i, \bm{x}_j)=\langle \varphi(\bm{x}_i), \varphi(\bm{x}_j)\rangle_{\mathcal{H}}.
	\label{eq:kernel_innerprod}
\end{equation}
A max-margin hyperplane will then be constructed in the feature space $\mathcal{H}$, corresponding to a potentially non-linear decision boundary in the input data space $\mathcal{X}$. Using a kernel function $k$ instead of an explicit feature map $\varphi$ allows us to access spaces where calculating an explicit feature map would be expensive or even impossible. The space $\mathcal{H}$ is hidden (or \emph{latent}) and only accessed indirectly.
In principle, a kernel allows classification of any kind of data, even data that is not numerical in nature (e.g. categorical data), as long as it can be mapped to some Hilbert space's inner product.

In QSVMs the Hilbert space of some quantum system is used as the latent space. The feature map encodes a datum $\bm{x_i}$ into a quantum state $\ket{\Psi(\bm{x_i})}$,
\begin{equation}	
	U(\bm{x_i}):\ketbra{0^n} \mapsto \ketbra{\Psi(\bm{x_i})},
\end{equation}
where $\ket{0^n} \coloneqq \ket{0}^{\otimes n} $. The resulting state depends non-linearly on $\bm{x_i}$~\cite{havlicek_supervised_2019, schuld_quantum_2019}.
\begin{figure}[h]
	\centering
	\begin{tikzpicture}
		\def\classicalZero{
			\begin{pmatrix}
				0 \\
				0 \\
				\vdots \\
				0
			\end{pmatrix}}
		\begin{yquant}
			qubit {} q[4];
			
			init {$\ket{0}$} (q);
			discard q[2];
			dots q[2];
			
			[x radius=0.7cm]
			box {$U(\bm{x_i})$} (q);
			hspace {3mm} q[0];
			[x radius=0.7cm]
			box {$U^{\dagger}(\bm{x_j})$} (q);
			measure q[0], q[1], q[3];
			
			dots q[2];
			output {$0^n$} (q);
		\end{yquant}
	\end{tikzpicture}
  \caption{Circuit realizing a quantum kernel. The probability of measuring the all-zero bitstring of length $n$, which is denoted by  ${0}^n$, is a similarity measure of $\ket{\Psi(\bm{x_i})}$ and $\ket{\Psi(\bm{x_j})}$.}
    \label{fig:quantum_kernel}
\end{figure}
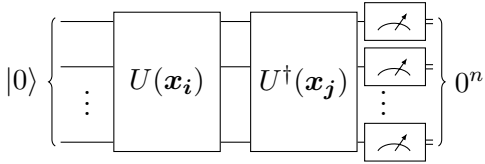
Now consider the circuit shown in Fig.~\ref{fig:quantum_kernel}. The probability $K_{ij}$ of measuring the all-zero bitstring $0^n$ is equivalent to the Hilbert-Schmidt inner product
\begin{align}
    K_{ij} =& \operatorname{Tr}\left[ \ketbra{\Psi(\bm{x_i})} \cdot \ketbra{\Psi(\bm{x_j})} \right]  \nonumber\\
    =& \left| \braket{\Psi(\bm{x_i})|\Psi(\bm{x_j})} \right|^2.
\end{align}
It is an inner product in a Hilbert space $\mathcal{H}$ and thus constitutes a valid kernel function
\begin{equation}
    k(\bm{x_i}, \bm{x_j}) = K_{ij} = \left| \braket{\Psi(\bm{x_i})|\Psi(\bm{x_j})} \right|^2.
\end{equation}
Since we cannot measure $K_{ij}$ directly, we must instead estimate it through many measurements (or \emph{shots}) $R$. This estimation is a stochastic process described by a binomial distribution and we find that finite-sampling bars us from evaluating the exact kernel. Instead, we only have access to a noisy version of the kernel
\begin{equation}
	\tilde{K}_{ij} = K_{ij} + \nu_{ij},
	\label{eq:noisy_kernel}
\end{equation}
where $\nu_{ij}$ describes the sampling noise with a variance
\begin{equation}
    \operatorname{Var}\left[ \nu_{ij} \right] = \frac{K_{ij} \cdot \left[1 - K_{ij}\right]}{R} = \mathcal{O}\left(\frac{1}{R}\right),
\end{equation}
and mean zero.

\section{Grover's algorithm}
Grover's algorithm~\cite{grover1996fast} is one of the most well-known quantum algorithms due to its simplicity and versatility. 
Grover's Algorithm solves the problem known as \emph{unstructured search}, where one needs to find one or multiple elements marked by a function $f$ among a set with no discernible structure present. One can think of this situation arising, for example, when filtering a database for entries with a certain value for a specific key, with $f$ being Boolean valued and 0 or \texttt{false} if the entry does not have the correct value and 1 or \texttt{true} if the value matches the search.

Classically, due to lack of structure, we can only evaluate $f$ on one element at a time and it takes on average $\frac{N}{2}$ queries to find the marked element. As a result, the classical algorithm has query complexity of $\mathcal{O}(N)$, scaling linearly with the size of the search space.

In the quantum case, we can however iteratively increase the amplitude of the marked states through an approach later generalized as amplitude amplification. This allows us to find the marked element with high probability in only $\mathcal{O}(\sqrt{N})$ queries.

\section{The classification problem}\label{sec:ProblemDescri}
The problem we want to solve revolves around the idea of classifying binary-valued \emph{functions} based on which elements they assign to 0 and which to 1.
Consider the space of the first $N=2^n$ non-negative integers $\mathbb{Z}_{N}$. We can define $N$ different indicator functions $\mathbb{I}_j:\mathbb{Z}_{N} \to \{0,1\}$ that mark a single element denoted by $j$,
\begin{equation}
	\mathbb{I}_j(t) =
	\begin{cases}
		1 & \text{if } j=t,\\
		0 & \text{otherwise}.
	\end{cases}
\end{equation}
For some $s\in \mathbb{Z}_{N/2}$ we assign classes to the functions $\mathbb{I}_j$ based on $j$ using the labeling rule or concept $y_s$,
\begin{equation}
	y_s(\mathbb{I}_j) =
	\begin{cases}
		+1 & \text{if } j \in \left[s, s+\frac{N - 2}{2}\right], \\
		-1 & \text{otherwise}.
	\end{cases}
\end{equation}
Note that for $s\in \mathbb{Z}_{N} \setminus \mathbb{Z}_{N/2} $ the classes are equivalent to the classes of $s'=s-N/2$ with swapped class labels. The classification problem is described in Fig.~\ref{fig:summary}(a) and visually represented in Fig.~\ref{fig:summary}(c).

The learning problem can be summarized as follows: The quantum or classical learner will be provided with a training set of $m=\text{poly}(n)$ tuples $\{(\mathbb{I}_j , y_s(\mathbb{I}_j)\}$ consisting of functions and their corresponding label. After training has been completed the learner has to classify unknown $\mathbb{I}_k$.

The learning problem we investigate demonstrates a separation in the oracular setting of Grover's algorithm, i.e., the training data is given as labeled black-box indicator functions. Towards practical applications, we discuss applying a related kernel to the real-world problem of pattern matching in App.~\ref{app:pattern}. This kernel in turn uses no oracles and input states are encoded as computational basis states. Although this approach makes weaker assumptions, we can still prove speedup in the worst-case complexity.
\section{Constructing a kernel}\label{sec:ConstructingKernel}
The classes contain functions that mark subsequent elements, therefore, we want to induce a bias into our kernel that assures that functions $\mathbb{I}_i$, $\mathbb{I}_j$ where $\norm{i-j}$ is small have high kernel values.
Consider to this effect a mapping to \emph{interval states}~\cite{liu_rigorous_2021}
\begin{equation}
	U(\mathbb{I}_j):\ket{0} \to \ket{\Psi(\mathbb{I}_j)}=\sqrt{\frac{1}{M}} \sum_{l=0}^{M-1} \ket{j + l},
	\label{eq:grover_interval_states}
\end{equation}
covering an interval of $2^m=M$ computational basis states. The construction of $U(\mathbb{I}_t)$ requires $n$ qubits and $n$ ancillas and can be split into four parts: First, we construct intervals in the first register by applying $m$ Hadamard gates,
\begin{equation}
	H^{\otimes m} \otimes \mathbbb{1}: \ket{0^n} \otimes \ket{0^n}
	\mapsto \left[\sqrt{\frac{1}{M}} \sum_l^{M-1} \ket{l}\right] \otimes \ket{0^n}.
	\label{eq:state_preparation_a}
\end{equation}
Then, using Grover's algorithm~\cite{grover_fast_1996} as a subroutine, we set the second register to the state corresponding to the marked element $i$. The success probability of Grover's algorithm exponentially converges to 1 in the size of the search space and is therefore assumed to be perfect in the following steps. 
We find, neglecting the normalization,
\begin{equation}
	\mathbbb{1} \otimes G_j:\sum_l^{M-1} \ket{l} \otimes \ket{0^n}
	\mapsto \sum_l^{M-1} \ket{l} \otimes \ket{j}.
	\label{eq:state_preparation_b}
\end{equation}
By using a quantum adder $\Sigma$ (see App.~\ref{app:adder}) we add the second register to the first,
\begin{equation}
	\Sigma: \sum_l^{M-1} \ket{l} \otimes \ket{j}
	\mapsto \sum_l^{M-1} \ket{j + l} \otimes \ket{j};
	\label{eq:state_preparation_c}
\end{equation}
and subsequently uncompute the second register containing the ancillas using Grover's algorithm:
\begin{equation}
	\mathbbb{1} \otimes G_j^\dagger:\sum_l^{M-1} \ket{j + l} \otimes \ket{j}
	\mapsto \sum_l^{M-1} \ket{j + l} \otimes \ket{0^n}.
	\label{eq:state_preparation_d}
\end{equation}
It is vital that we are able to uncompute the ancillae, otherwise the adjoint of the feature map will not produce the appropriate state. This is possible as the algorithm outputs a classical state (cf. App.\ref{ap:MME}).
\begin{figure}
	\centering
	\begin{tikzpicture}
		\begin{yquant}
			qubit {} q[4];
			qubit {} a[4];
			discard q[2];
			dots q[2];
			discard a[2];
			dots a[2];

			box {$H^{\otimes{m}}$} (q);
			box {Grover$(\mathbb{I}_j)$} (a);
			box {$\Sigma$ \\ 1$\mapsto$ 1+2} (q, a);
			box {Grover$^\dagger(\mathbb{I}_j)$} (a);
			dots q[2];
		\end{yquant}
	\end{tikzpicture}
	\caption{The feature map $U(\mathbb{I}_j)$ in circuit form. The system is initialized in $\ket{0^{2n}}$ The upper register will contain the interval state $\ket{\Psi(\mathbb{I}_j)}$, that is used in the kernel, whereas the lower register is made up of ancilla qubits that can be discarded.}
    \label{fig:grover_feature_map}
\end{figure}
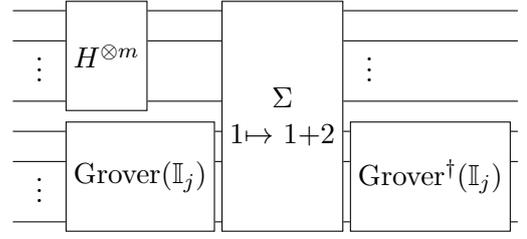
In Fig.~\ref{fig:grover_feature_map} the quantum circuit of the feature map is shown.
The quantum kernel (Fig.~\ref{fig:quantum_kernel}) implemented by this feature map is
\begin{equation}
	k(\mathbb{I}_i, \mathbb{I}_j)=
	\begin{cases}
		\frac{M-\abs{i - j}}{M} & \text{if } \abs{i - j} \leq M, \\
		0 & \text{otherwise}.
	\end{cases}
    \label{eq:interval_kernel}
\end{equation}
However,  we cannot measure the exact kernel value $K_{ij}$ due to finite sampling and need to work with the noisy estimate $\tilde{K}_{ij}$ instead (cf. \ref{sec:QSVM}). We can reduce the variance of $\tilde{K}_{ij}$ by running the circuit $R$ times since the variance $\nu_{ij}$ is upper-bounded by $1/R$. Naturally, the question arises how many shots $R$ are sufficient to ensure that the noise induced by finite sampling of the kernel does not strongly influence our classifier.
We map our problem to the \emph{noisy-halfspace learning problem} described in Ref.~\cite{liu_rigorous_2021} to prove robustness against finite sampling noise (see App.~\ref{app:nhlp}). In particular, this means that for training on $m=\operatorname{poly}(n)$ data points, $R \sim \mathcal{O}(m^4)$ shots are enough to bound 
the difference between the decision functions of classifiers trained with the exact kernel and noisy kernel below any desired constant with very high probability.

\section{Classical approach and query complexity}
We will now compare the QSVM to a classical machine learning classifier by first deriving the bounds on the query complexity of a classical learner.
We can easily construct a classical learner that uses a brute force approach to find the marked element of a given function. For a function $f:\mathbb{Z}_{N} \to \{0,1\}$ marking a single element, this problem is solved using a linear search which takes, on average, $N/2$ guesses.
A kernel such as Eq.~\eqref{eq:interval_kernel} is easy to calculate classically once the element has been found and
calculating a gram matrix for $m$ samples generally necessitates $\mathcal{O}(m^2)$ evaluations of the kernel. It follows that the query complexity of a classical kernel which sequentially searches for the marked element in the kernel is upper-bounded by $\mathcal{O}(d \hat{l} N)$ during classification of $\hat{l}$ new samples with $d$ support vectors.

Alternatively, the unstructured search can also be performed as a preprocessing step, meaning that it is not part of the kernel. One can then use the processed data to evaluate the kernel according to Eq.~\eqref{eq:interval_kernel} when needed. This reduces the complexity of classification to $\mathcal{O}(\hat{l} N)$ since we have already found the marked element for all support vectors $d$ during training. As shown in App.~\ref{app:classical_efficiency}, any classical classifier solving this problem is lower-bounded by $\mathcal{O}(N)$.

Now, let us consider the quantum SVM. The problem is then principally equivalent to the classical learner described above, with two major differences: First, we can identify the marked element using only $\mathcal{O}(\sqrt{N})$ queries, however we are working with an inexact, noisy kernel, requiring multiple evaluations $R$. Combining these results, we find that it takes
\begin{equation}
	\mathcal{O}\left(d \hat{l} R \sqrt{N} \right) = \mathcal{O}\left(d \hat{l} m^4 \sqrt{N} \right)
\end{equation}
evaluations to classify $\hat{l}$ new samples. It is important to note that $m$ scales with the problem size; in App.~\ref{app:nhlp},  it is shown that $m=\mathcal{O}(\text{poly}(n))$.

If we have access to a quantum computer we can also consider a hybrid approach: During preprocessing we use Grover's algorithm to find the marked elements for each $m$ training datapoints and $\hat{l}$ unseen datapoints, save the results and use them to evaluate the classical kernel. This reduces the complexity further to $\mathcal{O}(\hat{l} \sqrt{N})$. 
The preprocessing approach requires less computational resources since it only uses a single evaluation of Grover's algorithm compared to four for the fully quantum feature map described in Fig.~\ref{fig:grover_feature_map}.

In Fig.~\ref{fig:summary}(b) the query complexity of the different approaches is shown for the classification of one new sample; $\hat{l}=1$. The quantum approaches have a quadratic advantage over the classical learner. Limiting the quantum computation to a preprocessing step offers the best performance.

\section{Conclusion and outlook}
We have identified a learning problem where quantum kernel methods offer a provable quadratic advantage compared to classical machine learning. Crucial to such advantages is biasing the kernel in a problem-specific way~\cite{kuebler2021inductive}.

We arrived at our result using a feature map related to that described in Ref.~\cite{liu_rigorous_2021}, thereby demonstrating that their approach can be extended to other problems. 

Grover's algorithm serves as an instructive example of this flexibility. Other algorithms can be used in place of Grover's algorithm. To demonstrate the practical use of our proposed strategy we apply our kernel in a learning problem involving pattern matching in App.~\ref{app:pattern}, achieving a quadratic speedup. Algorithms with a larger advantage could 
provide a larger speedup in the context of a learning problem~\cite{childs_exponential_2003,kuperberg_subexponential-time_2005,hastings_short_2018,dalzell_mind_2022}.

Requiring the quantum subroutine to output a classical state, i.e., a basis state in a known basis, allows for the subsequent steps to be done classically. We show that the fully quantum kernel is inferior to a hybrid approach that utilizes a quantum computer for preprocessing, explicitly calculating the feature map, followed by a classical kernel calculation. This is in contrast with the original idea of the kernel trick where individual feature maps are hard (or impossible) to calculate directly, but the kernel function is computationally easy to evaluate. 

Our findings provide a new perspective on the origin of quantum advantages in machine learning algorithms, specifically algorithms of the structure of Ref.~\cite{liu_rigorous_2021}. If the quantum information processing can be shifted to a preprocessing step, thereby improving the performance of the algorithm, does it constitute a quantum advantage in learning \cite{gyurik_establishing_2022}? We provide an explicit specification of an algorithm, using quantum preprocessing, that improves on the fully quantum learner.
In practical problems one is less concerned whether a performance advantage or computational speedup emerges from the learning model itself or the data preprocessing steps. One can interpret the whole information processing pipeline, from the raw data preprocessing to the training of the statistical model, part of the learning algorithm. In real-world applications, the preprocessing steps, such as the normalization of the data, feature engineering and extraction~\cite{Li2018, Salau2019}, are crucial to the performance of the learner. Design choices made at the first steps of the pipeline can significantly hinder or enhance the performance of the learning model. 

Along these lines, there is an increased focus on what is known as data-centric AI within the classical AI community~\cite{motamedi2021, hajij2021, schmarje2022, zha2023}; for more details see Ref.~\cite{dcai21} and references therein. This paradigm emphasizes the importance of the data preprocessing procedure in stark contrast to the conventional approach that focuses on the architecture of the model, size of the dataset, and the amount of computing power.
There is much potential in the data-preprocessing steps: Only by adapting them in accordance with the structure of the data and the properties of the learner, the best performance can be achieved.
Our work extends the ideas above by demonstrating, for the first time to our knowledge, the potential of quantum computing to aid in these crucial preprocessing steps.
While the potential of quantum information processing for learning and for providing a  speed-up in computationally expensive subroutines of classical machine learning algorithms has been established, our research highlights the benefits that quantum computing can offer as part of the data preprocessing procedure.

\begin{acknowledgments}
We acknowledge discussion with Casper Gyurik, Vedran Dunjko and 
Ivan Rojkov. F.R. and E.Z. acknowledge funding from the Swiss National Science Foundation (Ambizione grant no. PZ00P2 186040) and the ETH Research Grant ETH-49 20-2. V.B. is supported by an ETH Research Grant (grant no. ETH C-04 21-2).

\end{acknowledgments}
\bibliography{ref_grover,Till_bib}{}
\bibliographystyle{quantum}

\onecolumn
\appendix
\newpage
\section{Application: Pattern matching}\label{app:pattern}
Pattern matching is a fundamental concept in computer science that refers to the process of searching for a specific pattern (i.e. a sequence of characters) within a larger piece of text. In the context of machine learning, pattern matching plays a critical role in tasks such as image and speech recognition and natural language processing~\cite{gu_recent_2018,ksaksamudre_review_2015,chowdhary_natural_2020}.

The problem of pattern matching can be described as such: Given a text $T$ of length $N$ and a search string (pattern) $P$ of length $L<N$ over a finite alphabet $\Sigma$. The task of the algorithm is to find an occurrence of $P$ in $T$ or to report that $P$ is not a substring. We will discuss the problem in one dimension however, it can be generalized to higher dimensions

The classical Knuth, Morris and Pratt  algorithm (Ref.~\cite{knuth_fast_1977}) solves the pattern matching problem in $\mathcal{O}(L+N)$ worst-case time complexity. This is optimal, as to be correct for all inputs a classical algorithm has to check every character of pattern and text. For random strings Ref.~\cite{yao_complexity_1979} proves a classical lower bound of $\Omega(\frac{N}{L} \log_q L)$ where $|\Sigma| = q$, the size of the alphabet.
The quantum algorithm presented in Ref.~\cite{ramesh_string_2000} provides a solution of the problem with worst-case query complexity $\mathcal{O}(\sqrt{N}+\sqrt{L})$ given oracular access to the text and pattern. For random patterns and texts quantum algorithms that are even faster on average are known~\cite{montanaro_quantum_2014}.

In the construction of our kernel we focus on the algorithm from Ref.~\cite{niroula_quantum_2021}. The data input is different from Ref.~\cite{ramesh_string_2000}: Pattern and text are provided as, classical, computational basis states rather than oracle queries. The algorithm worst-case time complexity $\mathcal{O}(\sqrt{N} ((\log(N))^2 +\log(L))$. It requires $\mathcal{O}(N+L)$ quantum space as it operates with classical data as an input to which it only has classical access. Here, we focus on optimizing time-complexity, so the space complexity is of no concern to us.

\subsection{Description of learning problem}
The learner is given a pattern $P$ of length $L$ and a set of texts $\{ T_t\}$ of length $N$ that are guaranteed to contain to contain the search string. The index $t$ denotes the (apriori unknown) location of the first bit of the pattern. We assign classes to the texts according to the position of the pattern $P$
\begin{equation}
    y_s(T_t) =	\begin{cases}
		+1 & \text{if } t \in \left[s, s+\frac{ \log(N) - 2}{2}\right], \\
		-1 & \text{otherwise}.
	\end{cases}
\end{equation}
The learner will be provided with a training set of $m = \text{poly}(\log(N))$ tuples $\{(T_j , y_s(T_j )\}$
consisting of functions and their corresponding
label. After training has been completed the
learner has to classify an unseen text $T_k$. We summarize the learning problem in Fig.~\ref{fig:summary}(a).

This application can be nicely motivated through tasks from image recognition. Instead of a one-dimensional string one can imagine a two-dimensional grid corresponding to pixels of an image. The classifier would then classify different images depending on where the searched pattern is located. 

\subsection{Summary of quantum algorithm for pattern matching}
Here, we briefly summarize the algorithm presented in Ref.~\cite{niroula_quantum_2021} for a more thorough analysis refer to the reference itself. 
We consider a tripartite quantum system of $n+N+L$ bits where $n=\log(N)$. The first register holds the index while second and third are used to store the text and pattern respectively. The text $T$ and the pattern $P$ are encoded in the computational basis states of the second and third registers,
\begin{align}
    \ket{\mathcal{T}} &= \ket{t_0t_1\dots t_{N-1}} = \bigotimes_{i=0}^{N-1} \ket{t_i},\\
    \ket{\mathcal{P}} &= \ket{p_0p_1\dots p_{L-1}} = \bigotimes_{j=0}^{L-1} \ket{p_j} .
\end{align}
For an alphabet of size 2 this can be done with identity and bitflip gates. We denote the encoding unitary as $E$.
After encoding the state of the system is
\begin{equation}
    \ket{\psi_1} = (\mathds{1}\otimes E) \ket{0^{ n+ N+L}} =\ket{0^{n }} \ket{\mathcal{T}} \ket{\mathcal{P}}.
\end{equation}
Assuming $N = 2^{n}$ we can apply a Walsh-Hadamard transform to the index register to generate a uniform superposition of computational basis states:
\begin{equation}
    \ket{\psi_2} = (H^{\otimes n} \otimes \mathds{1})\ket{\psi_1} = \left(\frac{1}{\sqrt{N}} \sum_{k=0}^{N-1} \ket{k} \right) \left( \bigotimes_{i=0}^{N-1} \ket{t_i} \right)\left( \bigotimes_{j=0}^{L-1} \ket{p_j}\right).
\end{equation}
Next, we apply the cyclic shift operator $S$ that shifts the qubits of the text register by $k$ sites according to the value of the index register $S\ket{k}\bigotimes_{i=0}^{N-1} \ket{t_i} =\ket{k}\bigotimes_{i=0}^{N-1} \ket{t_{i+k}}$. The implementation of the operator $S$ is detailed in Ref.~\cite{niroula_quantum_2021} and requires $\mathcal{O}(\log(N)^2)$ time steps. The displaced state is
\begin{equation}
     \ket{\psi_3} = ( S \otimes \mathds{1} )\ket{\psi_2}= \frac{1}{\sqrt{N}} \sum_{k=0}^{N-1} \left(\ket{k}  \bigotimes_{i=0}^{N-1} \ket{t_{(i+k)}} \right)\left( \bigotimes_{j=0}^{L-1} \ket{p_j}\right).
\end{equation}
In the following step, a XOR operation is performed between the first $L$ bits of the text register with the Pattern register. Such an operation can be performed unitarly using CNOT gates~\cite{zapusek_nonunitary_2023}. After this operation, we have
\begin{equation}
    \ket{\psi_4} = (\mathds{1}\otimes\text{CNOT} ^{\otimes L}) \ket{\psi_3}  =  \frac{1}{\sqrt{N}} \sum_{k=0}^{N-1} \left(\ket{k}  \bigotimes_{i=0}^{N-1} \ket{t_{(i+k)}} \bigotimes_{j=0}^{L-1} \ket{p_j \oplus t_{(j+k)}}\right).
\end{equation}
When the text that is displaced by the location of the pattern $t$ it is mapped to the all zero state in the pattern register. The index register in turn is in the state $t$.

Finally, $\mathcal{O}(\sqrt{N})$ steps of amplitude amplification are applied to the pattern register to amplify the amplitude of the $\ket{0^{L}}$ state. In total the algorithm requires $\mathcal{O}(\sqrt{N} ((\log(N))^2 +\log(L))$ time steps. The final state is
\begin{equation}
    \ket{\psi_\text{fin}} \approx  \ket{t} \left( \bigotimes_{i=0}^{N-1} \ket{t_{(i+t)}} \right)\left(\ket{0^{L}} \right).
\end{equation}
While overlap with states other than the desired output is $\mathcal{O}(\frac{1}{N})$. We will denote the action of the pattern matching algorithm on input text $T$ and pattern $P$ as $\text{Pattern}(T,P)$.

\subsection{Quantum kernel for pattern matching}
\begin{figure}
	\centering
	\begin{tikzpicture}
		\begin{yquant}
			qubit {} q[4];
			qubit {} a[4];
			init {$\ket{0^n}$} (q);
			init {$\ket{0^{n+N+L}}$} (a);
			discard q[2];
			dots q[2];
			discard a[2];
			dots a[2];
			
			box {$H^{\otimes{m}}$} (q);
			box {Pattern$(T_j,P)$} (a);
			box {$\Sigma$ \\ 1$\mapsto$ 1+2} (q, a[0,1]);
			box {Pattern$^\dagger(T_j,P)$} (a);
			dots q[2];
			output {$\ket{\Psi(T_j)}$} (q);
			output {$\ket{0^{n+N+L}} $} (a);
		\end{yquant}
	\end{tikzpicture}
	\caption{The feature map $U_P(T_j)$ in circuit form. The upper register will contain the interval state $\ket{\Psi(T_j)}$, that is used in the kernel, whereas the lower register is made up of ancilla qubits that can be discarded.}
 	\label{fig:pattern_feature_map}
\end{figure}
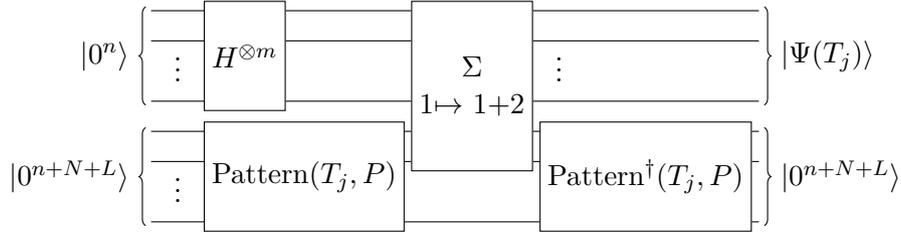

The feature map $U_P(T_j)$ encodes a text into a superposition of indices that follow the matching index of a given pattern $P$ in text $T_j$:
\begin{equation}\label{eq:featPattern}
    U_P(T_j) : \ket{0} \rightarrow \ket{\Psi(T_j)}= \sqrt{\frac{1}{M}} \sum_{l=0}^{M-1} \ket{j+l}.
\end{equation}
To implement the feature map $U_P(T_j)$, a quantum circuit is used that includes initialization of the index superposition, a pattern matching subroutine, a quantum adder, and uncomputing of the pattern matching, as depicted in Fig.~\ref{fig:pattern_feature_map}. This circuit is similar to the Grover feature map, but with the Grover subroutine replaced by pattern matching.
The feature map described in Eq.~\ref{eq:featPattern} implements a kernel
\begin{equation}
	k(T_i, T_j)=
	\begin{cases}
		\frac{M-\abs{i - j}}{M} & \text{if } \abs{i - j} \leq M, \\
		0 & \text{otherwise}.
	\end{cases}
    \label{eq:interval_kernel_pattern}
\end{equation}
This kernel returns a value between $0$ and $1$, where $1$ indicates a perfect match between the location of patterns in texts $T_i$ and $T_j$, and $0$ indicates a separation of more than $M$ indices. The kernel calculation is performed on a quantum computer by applying the feature map $U_P(T_i)$ and $U_P^\dag(T_j)$ to the initial state $\ket{0^n}$ and evaluating the probability of the all zero bitstring $0^n$ (Fig.~\ref{fig:quantum_kernel}).

In addition to text data, the feature map and kernel described in this work can be extended to other types of data, such as images. This is achieved by generalizing the pattern matching subroutine and interval states to multiple dimensions.
\subsection{Classical hardness}
The hardness of the problem depends on the input text. One can prove an advantage over a classical algorithm for pattern matching in the worst-case complexity. The worst-case complexity of classical pattern matching is  $\Theta(N+L)$~\cite{knuth_fast_1977}. Potentially, one has to check every  character of the pattern and text. The quantum algorithm from Ref.~\cite{niroula_quantum_2021} achieves the task in time $\mathcal{O}(\sqrt{N}(\log(N)^2+\log( L))$. It thereby provides a quadratic speedup.

The quantum classical separation for the learning problem can be proven analogously to that in App.~\ref{app:classical_efficiency}. The hardness in classification depends on the texts and is therefore only valid in worst-case complexity.
\section{Separability and the Noisy-halfspace Learning Problem}
\label{app:nhlp}
We will show, analogously to~\cite{liu_rigorous_2021}, that the mapping Eq.~\eqref{eq:grover_interval_states} produces a separable problem.
Define the \emph{halfspace state} of $s$
\begin{align}
	\ket{\Phi_s} = \sqrt{\frac{2}{N}} \sum_{i=0}^{\frac{N}{2}-1} \ket{s+i}
\end{align}
For ease of notation, we define the sets
\begin{align}
	c_s &= \left\{ \mathbb{I}_j | y_s(\mathbb{I}_j) = +1\right\}, \\
	\bar{c}_s &= \left\{\mathbb{I}_j | y_s(\mathbb{I}_j) = -1\right\},
\end{align}
corresponding to the set of functions in one class or the other.
Now consider a function $f$ fulfilling the following conditions:
\begin{enumerate}
	\item $f \in c_s$, i.e. $f(j)=\begin{cases}
		1 & \text{if } j=t,\\
		0 & \text{otherwise}.
	\end{cases}$ for some $t \in \left[s, s+\frac{N - 2}{2}\right]$
	\item $s+\frac{N -2}{2} -t \geq M$, therefore all computational basis states of the superposition $\ket{\Psi(f)}$ are also present in $\ket{\Phi_s}$.
\end{enumerate}
We define $\Delta$ as the overlap of the half-space state with $\ket{\Psi(f)}$;
\begin{align}
	\Delta =& \left| \braket{\Phi_s | \Psi(f)} \right|^2 = \frac{1}{M} \left| \sum_{l=0}^{M-1}  \underbrace{\braket{\Phi_s|t + l}}_{=\sqrt{2/N}} \right|^2 \nonumber \\
	=&\frac{1}{M} \frac{2}{N} M^2 = \frac{2M}{N} 
	\label{eq:def_delta}.
\end{align}
It is easy to see that all but $M$ functions in $c_s$ also fulfil the second condition. In particular, for a fraction of 
\begin{equation}
	1- \Delta = \frac{N/2 - M}{N/2}
\end{equation}
the second condition and thus equation Eq.~\eqref{eq:def_delta} hold.

With the equivalent calculation for functions $f \in \bar{c_s}$ we find
\begin{itemize}
	\item For a $1-\Delta$ fraction of $f \in c_s$, $\left| \braket{\Phi_s | \Psi(f)} \right|^2 = \Delta$,
	\item For a $1-\Delta$ fraction of $f \in \bar{c_s}$, $\left| \braket{\Phi_s | \Psi(f)} \right|^2 = 0$.
\end{itemize}
By choosing $\bm{w}_s=\ketbra{\Phi_s}$ and $b=-\frac{\Delta}{2}$ we find a suitable hyperplane that separates the data in the latent space $\mathcal{H}(N)$. In particular, the probability of a data point lying on the wrong side of the margin or within it is $\Delta$, the value of which we can essentially adjust freely by choice of a suitable $M$.

We see that this problem fulfills the conditions of the noisy halfspace learning problem defined in~\cite{liu_rigorous_2021}. From this it follows that for any $s$, training an SVM with the kernel $k$ using $R \sim m^4$ shots will return a classifier with accuracy of 0.99 or greater on a test dataset with probability $2/3$ over random training samples and noise.
Additionally, a number of $m=\text{poly}(n)$ training samples is sufficient for this purpose (see Ref.~\cite{liu_rigorous_2021} Eq. (D40)).

\section{Classical complexity}
\label{app:classical_efficiency}

Assume a classical learner is capable of determining the value $s$. During inference, the learner is provided with an unseen indicator function $f$. If the learner evaluates $f$ on $X$ values, corresponding to indicator functions of the same class, it finds the marked  element with probability $\frac{X}{N}$. Knowing $s$, it can then assign the correct class. With probability $1-\frac{X}{N}$ the correct marked element is not found, it then guesses the other class. This guess is correct with probability $P=\frac{1}{2}+\frac{X}{N}$. In total the classifier assigns the correct label with probability:
\begin{align}
    P_s = \frac{X}{N}+\left(\frac{1}{2}+\frac{X}{N}\right)\left(1-\frac{X}{N}\right)  \le \frac{1}{2}+\frac{3}{2}\frac{X}{N}.
\end{align}
To classify $f$ correctly with probability larger than some threshold (say $99\%$) we need that $X=\mathcal{O}(N)$.
\section{Quantum adder}\label{app:adder}

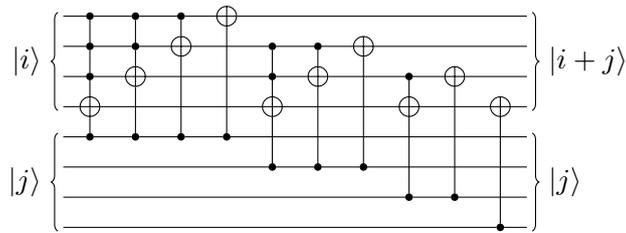
\begin{figure}
	\centering
	\begin{tikzpicture}
		\begin{yquant}
			qubit {} q[4];
			qubit {} a[4];
			init {$\ket{i}$} (q);
			init {$\ket{j}$} (a);
			
			cnot q[3] | q[0], q[1], q[2], a[0];
			cnot q[2] | q[0], q[1], a[0];
			cnot q[1] | q[0], a[0];
			cnot q[0] | a[0];
			
			cnot q[3] | q[1], q[2], a[1];
			cnot q[2] | q[1], a[1];
			cnot q[1] | a[1];
			
			cnot q[3] | q[2], a[2];
			cnot q[2] | a[2];
			
			cnot q[3] | a[3];
			output {$\ket{i + j}$} (q);
			output {$\ket{j}$} (a);
		\end{yquant}
	\end{tikzpicture}
	\caption{Quantum adder circuit for two 4 qubit registers. The lower register $\ket{j}$ is added to the upper register $\ket{i}$. Note that the uppermost qubit in each register corresponds to the lowest order bit.}
 	\label{fig:4qubit_quantum_adder}
\end{figure}

The quantum adder $\Sigma$ performs the action $\Sigma:\ket{i} \otimes \ket{j} \mapsto \ket{i+j} \otimes \ket{j}$ where $\ket{i}$ and $\ket{j}$ are computational basis states. A possible circuit that realizes such a unitary is pictured in Fig.~\ref{fig:4qubit_quantum_adder} for four qubits.

\section{On the extension to indicator functions that mark multiple elements}\label{ap:MME}

A natural extension to the Kernel and learning problem presented in the main text would be to extend it to indicator functions that mark multiple elements. However, for multiple reasons this extension fails. 

The Kernel presented in Sec.~\ref{sec:ConstructingKernel} relies on the fact that after we apply Grover's algorithm the state is in or close to a single state in a known basis. Specifically the computational basis. 
If one applies Grover to a function with multiple marked elements the output state is an even superposition of these elements. After applying the quantum adder (Fig.~\ref{fig:4qubit_quantum_adder}) the state of the ancilla system would be entangled with the first register. Due to this entanglement uncomputing the state on the ancilla register would also influence the main register. This would inhibit the proper function of the algorithm. 

A different approach would be to not use the adder entirely. Provided that the functions mark sufficient elements the superpositions output by Grover's algorithm would have enough overlap to ensure noise robustness. However, the indicator functions would have to mark so many elements that classically solving the problem becomes easier, thereby removing any quantum advantage.

\end{document}